# Deterministic quantum correlation between coherently paired photons acting on a beam splitter

Byoung S. Ham

School of Electrical Engineering and Computer Science, Gwangju Institute of Science and Technology

123 Chumdangwagi-ro, Buk-gu, Gwangju 61005, South Korea

(Submitted on Aug. 22, 2021; bham@gist.ac.kr)

**Abstract**

Quantum technologies based on the particle nature of a photon has been progressed over the last several decades, where the fundamental quantum features of entanglement have been tested by Hong-Ou-Mandel-type anticorrelation and Bell-type nonlocal correlation. Recently, mutually exclusive quantum features based on the wave nature of a photon have been investigated to understand the fundamental physics of 'mysterious' quantum correlation, resulting in deterministic and macroscopic quantum technologies. Here, we study the quantum natures of paired photons acting on a beam splitter, where mutual coherence plays a major role. Unlike current common understanding on anticorrelation, bipartite entanglement between paired photons does not have to be probabilistic or post-selected, but can be deterministic and even macroscopic via phase basis manipulation without violating quantum mechanics.

**Introduction**

A corpuscular nature of light has been accepted since the photoelectric effect was observed by Einstein in 1905. Entanglement [1] between paired photons has been understood in terms of the particle nature of a photon, where a photon, on the other hand, also has a wave nature governed by Maxwell's wave equations [2-4]. Based on the fundamental understanding of complementarity theory or wave-particle duality in quantum physics [2], these two mutually exclusive natures of a photon are complementary [5,6]. Thus, one nature (energy) of a photon cannot simultaneously appear alongside the other (phase). In other words, specifying a photon's energy in a Fock state results in vagueness for its phase information. This fundamental physics of the energy-phase uncertainty relationship, however, does not apply to paired photons. Without understanding of the phase basis in such a coupled system, quantum features become mysterious as well as probabilistic. In that sense, clear definitions of the classicality and quantumness of the paired system are prerequisite to discuss quantum correlation. The understanding on quantum correlations between two or more bipartite photons has recently been revisited to account for the wave nature with a mutual coherence basis, even though individual photons have no specific phase information [7,8]. Here, we investigate a paired photon system that acts on a beam splitter (BS) to understand the fundamental quantum characteristics of a nonclassical feature.

Entanglement between paired photons or atoms is known as a weird quantum phenomenon, where specifying the mutual phase relationship has not been considered [9-20]. Thus, post-measurement techniques have been developed for the probabilistic nature of quantum correlation. Recently, a completely different approach has been applied for the fundamental physics of anticorrelation [7], photonic de Broglie waves [8], and Franson-type nonlocal correlation [21]. In regard to quantum resources of light, spontaneous parametric down conversion (SPDC)-based $\chi^{(2)}$ nonlinear optical processes have been applied for entangled photon-pair generations [10,11]. Due to the vagueness of phase information between SPDC-generated entangled photons, however, to date post-measurement-based quantum features have been nondeterministic and even misleading [22,23]. Here, coherent photon pairs acting on a BS are investigated for the same quantum feature of anticorrelation observed in a Hong-Ou-Mandel (HOM) dip. Unlike the common understanding of the probabilistic nature of a photon on a BS, mutual coherence between paired photons results in a definite output relationship regardless of measurements [7,23]. Based on the correct understanding of quantum features by mutually coherent photon pairs, the conventional particle nature-based non-determinacy is compared to access a critical mismatch between them. As a result, the 'mysterious' quantum features of entanglement are determined to be deterministic and macroscopic.

**Analysis**

Figure 1 shows the most fundamental quantum measurement scheme for paired photons that act on a BS [10]. The paired photons are entangled by the SPDC processes [24]. For the SPDC-entangled photon pairs, a standard particle nature-based analysis cannot result in a deterministic quantum feature because a definite phase relationship between the paired photons is neglected [10,11,14-19]. As already observed by many research groups over the last several decades, the anticorrelation or a HOM dip [10] indicates entanglement between the paired photons, where anticorrealtion results from a photon bunching phenomenon in the output ports. The photon bunching is caused by destructive quantum interference between the paired photons, and requires a predetermined phase relationship between them [7,23]. This mutual phase relationship between the paired photons does not violate quantum mechanics. Although each photon's phase information cannot be determined by quantum mechanics, specifying a relative phase between the paired photons is permitted as discussed by Einstein, Podolsky, and Rosen (EPR) [1] and Bohm [2] with spin polarization basis.

One of the most fundamental quantum features is randomness, which results in quantum superposition between measurement bases. For a typical Young's double slit or an Mach Zehnder interferometer (MZI), the fundamental phase basis set is $\theta \in \{0, \pi\}$, where the basis relates to a pure state. In Fig. 1(a), the fundamental phase basis set of the entangled photon system for a BS is also $\varphi \in \{0, \pi\}$, resulting in photon bunching in the output ports [7]. Here, the randomness on a BS is for individual photons, not for the paired photon, resulting in self-interference in an MZI [25], where the measurement probability in the output ports is equal: $I_A = I_B = \frac{I_0}{2}$ and $I_j = E_j E_j^*$. The probability amplitude of $E_j$ for each photon is the fundamental resource of the randomness known as Born's rule [26-28]. This probability amplitude is the fundamental difference from classical physics. However, the fundamental physics of the quantum feature of anticorrelation cannot be completely understood unless the mutual phase relation is involved for a coupled system, because otherwise the random intensity distribution would be the same as each photon prohibiting anticorrelation [7,23].

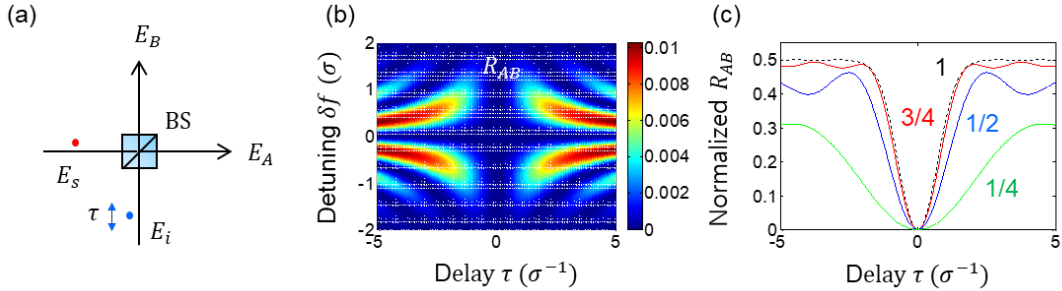

Fig. 1. SPDC generated photon pair interactions on a beam splitter for anticorrelation. (a) Schematic of bipartite photon pair interactions. (b)-(c) Numerical calculations for equation (3) with $\delta\varphi' = \pm\pi/2$. In (c), the detuning $\delta f$ values are averaged and normalized with respect to the delay $\tau$. The numerical values indicate the ratio of reduced bandwidth to the original in (b): dotted (1), red (3/4), blue (1/2), green (1/4). BS: a 50/50 nonpolarizing beam splitter. The entangled photon distribution is Gaussian with standard deviation $\sigma$.

For the paired input photons in Fig. 1(a), each photon's amplitude can be described as $E_j = E_0 e^{i\varphi_j}$, where $\varphi_j$ of each photon j is completely random and nondeterministic. The time delay $\tau$ between $E_s$ and $E_i$ is with respect to the arrival on the BS. Using the BS matrix representation [29], the following outputs are obtained for Fig. 1(a):

$$I_A = I_0(1 - \sin\delta\varphi), \qquad (1)$$
$$I_B = I_0(1 + \sin\delta\varphi), \qquad (2)$$

where $E_A = \frac{1}{\sqrt{2}}(E_s + iE_i)$, $E_B = \frac{i}{\sqrt{2}}(E_s - iE_i)$, $I_0 = E_0 E_0^*$, and $\delta\varphi = \varphi_s(\delta, \tau) - \varphi_i(\delta, \tau) + \delta\varphi'$. Here, $\delta\varphi'$ is the initially given mutual phase difference between the paired photons generated by the SPDC process without violation of quantum mechanics. Thus, equations (1) and (2) are basically the same as the Young's double slit system governed by coherence optics if $\delta\varphi'$ is added. The normalized coincidence measurement between $I_A$ and $I_B$ is as follows:

$$R_{AB}(\tau) = \cos^2(\delta\varphi). \qquad (3)$$

To satisfy the observed anticorrelation in a HOM dip at $\tau = 0$, i.e. $R_{AB}(\tau = 0) = 0$ [10], the initially given phase difference must be $\delta\varphi' = \pm\pi/2$, resulting in $R_{AB} = sin^2(\varphi)$ and $\varphi = \varphi_s(\delta,\tau) - \varphi_i(\delta,\tau)$ [7]. For the relative delay $\tau$ between the paired input photons, the relative phase $\varphi$ becomes nonzero due to the frequency difference $\delta f(\tau)$ in a given bandwidth, as shown in Fig. 1(b).

The coincidence measurements resulting from an ensemble average for all $\delta f$ at a given time delay $\tau$ are shown in Figs. 1(b) and (c). Thus, the original zero coincidence detection at $\tau = 0$ is alleviated as $\tau$ increases, as shown in Fig. 1(c) (see the dotted curve for full bandwidth). This $\tau$-dependent quantum correlation loss is maximized if $\tau \geq 2\sigma^{-1}$, where $\sigma$ is the standard deviation of the photon spectral distribution. Here, $R_{AB} = 1/2$ represents the classical lower bound applied to incoherent photons [7]. However, for the spectrally narrowed input photons, such correlation loss is lessened as shown by colored curves in Fig. 1(c) due to enhanced coherence with respect to the $\tau$-dependent $\delta f$. The reason for the lack of λ-dependent fringe observed in HOM dips is due to the $\tau$-resulting coherence washout in an ensemble, as theoretically demonstrated [23]. Thus, the anticorrelation on a BS for a HOM dip is no longer mysterious, and is instead deterministic based on the relative phase between paired photons. In other words, there is no possibility of equal intensity $(I_A = I_B)$ at $\tau = 0$ for the SPDC-entangled photon pairs that act on a BS due to the predetermined phase relationship.

Figure 2 shows corresponding scenarios of the coherent photon pair-based measurements for photon bunching on the first BS, where Fig. 1(a) is modified with an additional set of BS. In Fig. 2, we investigate the same quantum features of anticorrelation based on the coherent photon pairs. This analysis leads to the same entanglement as observed by SPDC photon sources in Fig. 1. For context, Bell's inequality [9] regarding the EPR paradox [1] indicates that quantum correlation violates the inequality relationship of the classical correlation between incoherently paired particles. As is already known, the quantum correlation between paired entities is the key feature of the Bell's inequality violation. This nonlocal quantum correlation is based on a specific phase relationship between the paired photons. Further, the term of classicality must be defined for individual corpuscles like identical beans or stones without any phase relationship between them. Coherent photons from an attenuated laser may be accessed with the classical feature if they do not satisfy sub-Poisson distribution. However, with conditional measurements, an attenuated laser can satisfy sub-Poisson photon characteristic, because the majority of vacuum states can be eliminated and only doubly bunched photons can be conditionally selected. This quantum feature of coherent photons has been experimentally demonstrated by a coincidence detection technique [30] and can also be accomplished in Fig. 2 for photon resolving.

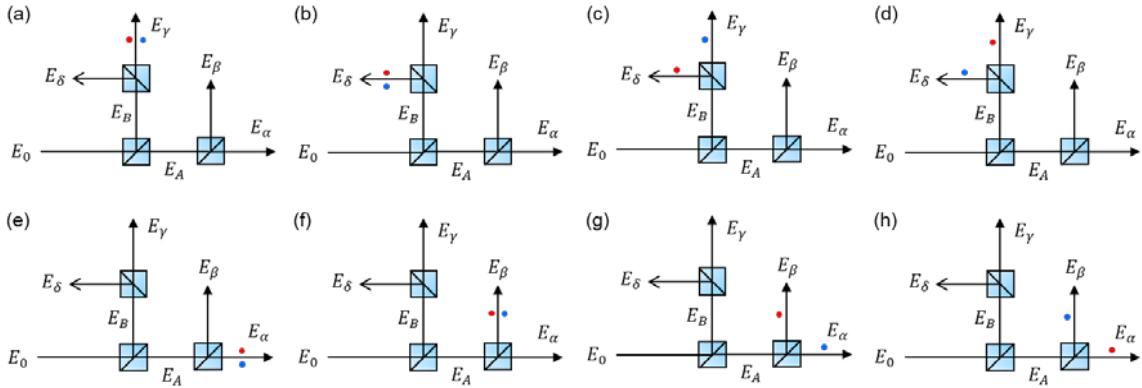

Fig. 2. Photon bunching scenarios on a BS for paired coherent input photons. The red and blue dots indicate identical and coherent photons. BS: 50/50 nonpolarizsing beams splitter.

In Fig. 2, coherently paired doubly bunched photons impinge on a BS along the same horizontal direction as denoted by $E_0$. Even if the single photon detectors do not resolve the photon number, the cascade BS scheme represented in Fig. 2 can be applied as a good test tool to determine whether the coherently paired input photons are bunched through the BS. The statistical error of three or more bunched photon cases can be eliminated according to Poisson statistics with adequate selection of a mean photon number (see section A of the Supplementary Information) [30]. If bunched photons result in one output port of $E_A$ or $E_B$, this would indicate four different single photon detections of $E_\alpha$, $E_\beta$, $E_\gamma$, and $E_\delta$, as shown in Fig. 2. Non-zero coincidence measurements between $E_\alpha$ and $E_\beta$ ($E_\gamma$ and $E_\delta$) clearly demonstrate the existence of bunched photons in $E_A$ ($E_B$). On the contrary, the total number of non-bunched photon cases on the first BS is also eight, and results in zero coincidence detection between $E_\alpha$ and $E_\beta$ ($E_\gamma$ and $E_\delta$) (see section B of the Supplementary Information).

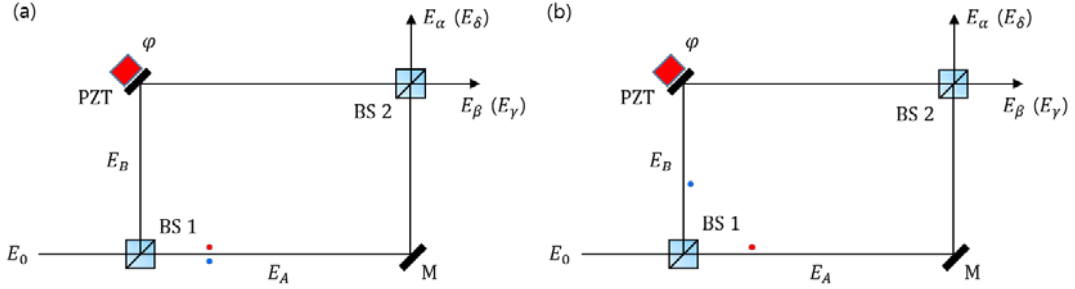

Fig. 3. Possible photon characteristics in an MZI. (b) Photon bunching. (b) Photon antibunching. BS: 50/50 nonpolarizing beam splitter, M: mirror, PZT: piezo-electric transducer. The red and blue dots indicate coherent photons.

For the analysis of the bunched photon cases in Fig. 2, the original single BS scheme is modified for an MZI model in Fig. 3 with the relative phase control of $\varphi$ ($= \varphi_{AB}$), where $\varphi_{AB} = \varphi_B - \varphi_A$. Figure 3(a) is for the case of photon bunching on BS 1, whereas Fig. 3(b) is for antibunching. Using matrix representation of a BS and a phase shifter [29], the following output fields are obtained:

$$\begin{bmatrix} E_\alpha \\ E_\beta \end{bmatrix} = \frac{1}{2}\begin{bmatrix} 1 & 0 \\ 0 & e^{i\varphi_{\alpha\beta}} \end{bmatrix}\begin{bmatrix} 1 & i \\ i & 1 \end{bmatrix}\begin{bmatrix} 1 & 0 \\ 0 & e^{i\varphi} \end{bmatrix}\begin{bmatrix} 1 & i \\ i & 1 \end{bmatrix}\begin{bmatrix} E_o \\ 0 \end{bmatrix}$$
$$= \frac{E_o}{2}\begin{bmatrix} 1 & 0 \\ 0 & e^{i\varphi_{\alpha\beta}} \end{bmatrix}\begin{bmatrix} 1-e^{i\varphi} & i(1+e^{i\varphi}) \\ i(1+e^{i\varphi}) & -(1-e^{i\varphi}) \end{bmatrix}\begin{bmatrix} 1 \\ 0 \end{bmatrix}, \quad (4)$$

where $\varphi_{\alpha\beta} = \varphi_\beta - \varphi_\alpha$ is between $E_\alpha$ and $E_\beta$. Thus, $E_\alpha = \frac{E_o}{2}(1 - e^{i\varphi})$ and $E_\beta = \frac{iE_o}{2}e^{i\varphi_{\alpha\beta}}(1 + e^{i\varphi})$. Finally, the corresponding intensities are $I_\alpha = \frac{I_o}{2}(1 - cos\varphi)$ and $I_\beta = \frac{I_o}{2}(1 + cos\varphi)$, as observed by single photons [25]. This means that photon bunching either for $E_A$ or $E_B$ in Fig. 3(a) is completely prohibited for the phase basis $\theta \in \{0, \pi\}$, otherwise $I_\alpha$ and $I_\beta$ cannot be zero due to the same random-path selection by each photon. This fact is known as Born's rule for quantum measurements [26-28], where single photon-based self-interference is due to the superposition of both paths ($E_A$ and $E_B$) in an MZI and thus prohibits any kind of photon bunching phenomenon on a BS, resulting in Fig. 3(b) with $I_A = I_B = I_0/2$. Therefore, potential scenarios of Fig. 2 for photon bunching on a BS contradict quantum mechanics. Here, Fig. 3(b) is exactly the same as Fig. 1(a) for anticorrelation on a BS at $\varphi = 0$, except for the cancellation of coherence washout [7].

Born's rule on measurements for a single photon in quantum mechanics intuitively gives the same answer to the coherent photons, where both single photon and coherent light behave in the same way in an interferometric system, in which the Sokin parameter is limited to the two-input, two-output system of a BS [26]. However, the phase controllability of paired coherent photons has been experimentally demonstrated for the same anticorrelation [30], which is also contradictory to our current common understanding on probabilistic nature of single photons. Thus, the wave nature of a photon results in a clear understanding on the fundamental physics of quantum correlation on a BS. The so-called 'mysterious' quantum phenomenon has now been clarified and found to be deterministic. Further, the quantum determinacy in a coupled system does not violate quantum mechanics.

**Conclusion**
Photon characteristics on a BS were investigated for both SPDC-generated entangled photon pairs and coherently provided photon pairs. For the entangled photon pairs, the BS matrix representation based on the wave nature of a photon results in a definite relative phase difference between the paired photons. Based on the relative time delay, numerical simulations for coincidence measurements between two output ports resulted in bandwidth-dependent coherence washout due to frequency detuning-dependent random phases on the averaging process. If the photon bandwidth is spectrally reduced, such coherence washout was alleviated, resulting in modulation fringe, as previously observed [31] and analyzed [23].

For coherently bunched photons as an input, the output photon distribution from a BS was contradictory to our common understanding of random choice of output ports as in independent and incoherent single photon cases. Such a destructive interference on a BS for photon bunching was analyzed in an MZI scheme and concluded not to be possible for coherently paired photons having the same input direction. Unlike common understanding on coherence optics, coherently paired photons can also contribute to quantum correlation as observed recently [30]. Thus, the relative phase information between paired photons is a definite control parameter for deterministic quantum correlation in an interferometric system. Due to the wave nature of a photon, macroscopic quantum correlation is inherent by collective manipulation of phase basis information of the photon-BS system.

**Acknowledgment**
BSH acknowledges that this paper was motivated by Prof. J. Lee of Hanyang University, S. Korea for coherent photon bunching on a BS. This work was supported by Gwangju Institute of Science Technology via GRI2021.